\documentclass[prl,reprint,nofootinbib,altaffilletter,superscriptaddress,showpacs]{revtex4-1}
\usepackage{amsmath,amssymb,amsfonts}
\usepackage[dvips]{graphicx}
\usepackage{dcolumn}
\usepackage{bm}
\usepackage{bbm}
\usepackage{xcolor}
\usepackage{hyperref}
\usepackage{lineno}
\usepackage{ulem}

\def\figwidth{0.49\textwidth}
\begin{document}


\title{$\omega N$ scattering length from $\omega$ photoproduction on the proton near the threshold}
\author{T.~Ishikawa}
\email[Corresponding author: ]{ishikawa@lns.tohoku.ac.jp}
\affiliation{Research Center for Electron Photon Science (ELPH), Tohoku University, Sendai, Miyagi 982-0826, Japan}
\author{H.~Fujimura}
\altaffiliation[Present address: ]{Department of Physics, Wakayama Medical University, Wakayama 641-8509, Japan}
\affiliation{Research Center for Electron Photon Science (ELPH), 
Tohoku University, Sendai, Miyagi 982-0826, Japan}
\author{H.~Fukasawa}
\affiliation{Research Center for Electron Photon Science (ELPH), 
Tohoku University, Sendai, Miyagi 982-0826, Japan}
\author{R.~Hashimoto}
\altaffiliation[Present address: ]{Institute of Materials Structure Science (IMSS), KEK, Tsukuba 305-0801, Japan}
\affiliation{Research Center for Electron Photon Science (ELPH), 
Tohoku University, Sendai, Miyagi 982-0826, Japan}
\author{Q.~He}
\altaffiliation[Present address: ]{Department of Nuclear Science and Engineering, Nanjing University of Aeronautics and Astronautics (NUAA), Nanjing 210016, China}
\affiliation{Research Center for Electron Photon Science (ELPH), 
Tohoku University, Sendai, Miyagi 982-0826, Japan}
\author{Y.~Honda}
\affiliation{Research Center for Electron Photon Science (ELPH), 
Tohoku University, Sendai, Miyagi 982-0826, Japan}
\author{A.~Hosaka}
\affiliation{Research Center for Nuclear Physics (RCNP), Osaka University, Ibaraki 567-0047, Japan}
\affiliation{
Advanced Science Research Center, Japan Atomic Energy Agency (JAEA), Tokai 319-1195, Japan}
\author{T.~Iwata}
\affiliation{Department of Physics, Yamagata University, Yamagata 990-8560, Japan}
\author{S.~Kaida}
\affiliation{Research Center for Electron Photon Science (ELPH), 
Tohoku University, Sendai, Miyagi 982-0826, Japan}
\author{J.~Kasagi}
\affiliation{Research Center for Electron Photon Science (ELPH), 
Tohoku University, Sendai, Miyagi 982-0826, Japan}
\author{A.~Kawano}
\affiliation{Department of Information Science, Tohoku Gakuin University, Sendai 981-3193, Japan}
\author{S.~Kuwasaki}
\affiliation{Research Center for Electron Photon Science (ELPH), 
Tohoku University, Sendai, Miyagi 982-0826, Japan}
\author{K.~Maeda}
\affiliation{Department of Physics, Tohoku University, Sendai 980-8578, Japan}\author{S.~Masumoto}
\affiliation{Department of Physics, University of Tokyo, Tokyo 113-0033, Japan}\author{M.~Miyabe}
\affiliation{Research Center for Electron Photon Science (ELPH), 
Tohoku University, Sendai, Miyagi 982-0826, Japan}
\author{F.~Miyahara}
\altaffiliation[Present address: ]{Accelerator Laboratory, KEK, Tsukuba 305-0801, Japan}
\affiliation{Research Center for Electron Photon Science (ELPH), 
Tohoku University, Sendai, Miyagi 982-0826, Japan}
\author{K.~Mochizuki}
\affiliation{Research Center for Electron Photon Science (ELPH), 
Tohoku University, Sendai, Miyagi 982-0826, Japan}
\author{N.~Muramatsu}
\affiliation{Research Center for Electron Photon Science (ELPH), 
Tohoku University, Sendai, Miyagi 982-0826, Japan}
\author{A.~Nakamura}
\affiliation{Research Center for Electron Photon Science (ELPH), 
Tohoku University, Sendai, Miyagi 982-0826, Japan}
\author{S.X.~Nakamura}
\affiliation{University of Science and Technology of China, Hefei 230026, China}
\author{K.~Nawa}
\affiliation{Research Center for Electron Photon Science (ELPH), 
Tohoku University, Sendai, Miyagi 982-0826, Japan}
\author{S.~Ogushi}
\affiliation{Research Center for Electron Photon Science (ELPH), 
Tohoku University, Sendai, Miyagi 982-0826, Japan}
\author{Y.~Okada}
\affiliation{Research Center for Electron Photon Science (ELPH), 
Tohoku University, Sendai, Miyagi 982-0826, Japan}
\author{K.~Okamura}
\affiliation{Research Center for Electron Photon Science (ELPH), 
Tohoku University, Sendai, Miyagi 982-0826, Japan}
\author{Y.~Onodera}
\affiliation{Research Center for Electron Photon Science (ELPH), 
Tohoku University, Sendai, Miyagi 982-0826, Japan}
\author{K.~Ozawa}
\affiliation{Institute of Particle and Nuclear Studies, High Energy Accelerator Research Organization (KEK), Tsukuba 305-0801, Japan}
\author{Y.~Sakamoto}
\affiliation{Department of Information Science, Tohoku Gakuin University, Sendai 981-3193, Japan}\author{M.~Sato}
\affiliation{Research Center for Electron Photon Science (ELPH), 
Tohoku University, Sendai, Miyagi 982-0826, Japan}
\author{T.~Sato}
\affiliation{Research Center for Nuclear Physics (RCNP), Osaka University, Ibaraki 567-0047, Japan}
\author{H.~Shimizu}
\affiliation{Research Center for Electron Photon Science (ELPH), 
Tohoku University, Sendai, Miyagi 982-0826, Japan}
\author{H.~Sugai}
\altaffiliation[Present address: ]{Gunma University Initiative for Advanced Research (GIAR), Maebashi 371-8511, Japan}
\affiliation{Research Center for Electron Photon Science (ELPH), 
Tohoku University, Sendai, Miyagi 982-0826, Japan}
\author{K.~Suzuki}
\altaffiliation[Present address: ]{The Wakasa Wan Energy Research Center, Tsuruga 914-0192, Japan}
\affiliation{Research Center for Electron Photon Science (ELPH), 
Tohoku University, Sendai, Miyagi 982-0826, Japan}
\author{Y.~Tajima}
\affiliation{Department of Physics, Yamagata University, Yamagata 990-8560, Japan}
\author{S.~Takahashi}
\affiliation{Research Center for Electron Photon Science (ELPH), 
Tohoku University, Sendai, Miyagi 982-0826, Japan}
\author{Y.~Taniguchi}
\affiliation{Research Center for Electron Photon Science (ELPH), 
Tohoku University, Sendai, Miyagi 982-0826, Japan}
\author{Y.~Tsuchikawa}
\altaffiliation[Present address: ]{Department of Physics, Nagoya University, Nagoya 464-8602, Japan}
\affiliation{Research Center for Electron Photon Science (ELPH), 
Tohoku University, Sendai, Miyagi 982-0826, Japan}
\author{H.~Yamazaki}
\altaffiliation[Present address: ]{Radiation Science Center, KEK, Tokai 319-1195, Japan}
\affiliation{Research Center for Electron Photon Science (ELPH), 
Tohoku University, Sendai, Miyagi 982-0826, Japan}
\author{R.~Yamazaki}
\affiliation{Research Center for Electron Photon Science (ELPH), 
Tohoku University, Sendai, Miyagi 982-0826, Japan}
\author{H.Y.~Yoshida}
\affiliation{Department of Physics, Yamagata University, Yamagata 990-8560, Japan}
\begin{abstract}
Photoproduction of the $\omega$ meson on the proton 
has been experimentally studied 
near the threshold.
The total cross sections are
determined at incident energies ranging from 1.09 to 1.15 GeV.
The 1/2 and 3/2 spin-averaged scattering length
$a_{\omega p}$ and effective range $r_{\omega p}$ 
between the $\omega$ meson and proton are
estimated from the shape of the total cross section 
as a function of the incident photon energy:
$a_{\omega p} = \left(-0.97^{+0.16_{\rm stat}}_{-0.16_{\rm stat}}{}^{+0.03_{\rm syst}}_{-0.00_{\rm syst}}\right)+
i \left(0.07^{+0.15_{\rm stat}}_{-0.14_{\rm stat}}{}^{+0.17_{\rm syst}}_{-0.09_{\rm syst}}\right)$
fm and $r_{\omega p}=\left(+2.78^{+0.68_{\rm stat}}_{-0.54_{\rm stat}}{}^{+0.11_{\rm syst}}_{-0.13_{\rm syst}}
\right)+i\left(-0.01^{+0.46_{\rm stat}}_{-0.50_{\rm stat}}{}^{+0.07_{\rm syst}}_{-0.00_{\rm syst}}\right)$ fm,
resulting in a repulsive force.
The real and imaginary parts for $a_{\omega p}$ and $r_{\omega p}$
are determined separately for the first time.
A small $P$-wave contribution does not affect the obtained values.
\end{abstract}


\maketitle
The structure of hadrons and dynamical hadron-mass generation
are the most important subjects to be studied 
in the non-perturbative domain of quantum chromodynamics (QCD).
The $\omega$ meson ($\omega$)  is one of the best established 
hadrons, and it is considered to give a short-ranged repulsive central force
and a strong spin-orbit force between two nucleons ($N$s)~\cite{machleidt}.
Nevertheless, 
the fundamental properties of $\omega$ such as the interaction with $N$
is not known yet due to the difficulties in realizing scattering experiments.
Detailed information on $\omega N$ scattering would not 
only reveal highly excited nucleon resonances ($N^*$)
but also 
have a strong relevance to the equation of state (EoS) describing 
the interior of neutron stars~\cite{sxn1}.
Gravitational wave observations just have begun to provide information 
on EoS~\cite{sxn2}. 

The low-energy $\omega N$ scattering is characterized 
by the scattering length $a_{\omega N}$ and effective range $r_{\omega N}$
through an effective-range expansion of the $S$-wave phase shift $\delta(p)$:
\begin{equation}
p \cot\delta(p) = \frac{1}{a_{\omega N}} + \frac{1}{2} \, r_{\omega N}\, p^2 + O(p^4),
\end{equation}
where $p$ denotes the momentum of $\omega$
in the $\omega N$ center-of-mass (CM) frame.
A positive (negative) Re $a_{\omega N}$ gives 
attraction (repulsion), 
and 
a positive Im $a_{\omega N}$
corresponds to the absorption
to another channel such as $\omega N \to \pi N$. The $r_{\omega N}$ 
provides the momentum dependence of the interaction.
Recently, the A2 collaboration at the Mainz MAMI facility
has reported $\left|a_{\omega N}\right|=0.82\pm 0.03$ fm,
which is extracted from $\omega$ photoproduction on the proton 
($\gamma p \to \omega p$) 
near the threshold assuming a vector meson dominance (VMD) model~\cite{mainz}.
The obtained value is a combination of two 
independent $S$-wave scattering lengths with total spins
of 1/2 and 3/2. The unknown sign of $a_{\omega N}$ leaves the 
naive question of whether low-energy $\omega N$ scattering 
is repulsive or attractive. 

Theoretically
 estimated values of $a_{\omega N}$ are scattered 
in a wide range from attractive to repulsive ones.
The effective Lagrangian approach based on chiral symmetry 
gives an attractive value of $a_{\omega N}=+1.6+i0.30$ fm~\cite{omega1}.
A QCD sum-rule analysis provides a weakly attractive value of 
$a_{\omega N}=+0.41\pm 0.05$ fm~\cite{omega2}.
The coupled-channel unitary approach gives repulsive values of
$a_{\omega N}^{(1/2)}=-0.45+i0.31$ fm and 
$a_{\omega N}^{(3/2)}=-0.43+i0.15$ fm for the two total spins, giving 
a spin-averaged value of $a_{\omega N}=-0.44+i0.20$ fm~\cite{omega3}.
The coupled-channel analysis of $\omega$ production in pion
and photo-induced reactions gives a very weakly 
repulsive value of $a_{\omega N} = -0.026+i0.28$ fm~\cite{omega4}.
The dynamical coupled-channel analysis resulted in
$a_{\omega N}^{(1/2)}=0.0454+i0.0695$ fm and 
$a_{\omega N}^{(3/2)}=-0.180+i0.0597$ fm,
giving a repulsive spin-averaged value\footnote{
We adopt $a_{\omega N} = (1/3) a_{\omega N}^{(1/2)}
+ (2/3) a_{\omega N}^{(3/2)}$ for the spin average using the convention of 
Lutz {\it et al.}~\cite{omega3,omega4}.} of 
 $a_{\omega N}=-0.135+i0.0630$ fm~\cite{omega5}.
Neither
the coupled-channel analyses nor the VMD
analysis by the A2 collaboration 
incorporates the finite width of $\omega$ in the final state.

To determine the low-energy $\omega N$ scattering parameters 
$a_{\omega N}$ and $r_{\omega N}$ experimentally,
we investigate the $\gamma p \to \omega p$ reaction
very close to the reaction threshold.
Several collaborations have already measured the total cross sections 
near the threshold using the $\omega\to \pi^+\pi^-\pi^0$ decay mode 
(SAPHIR~\cite{saphir} and CLAS~\cite{clas} collaborations),
and the $\omega \to \pi^0\gamma$ decay mode
(CBELSA/TAPS~\cite{cbelsa} 
and A2~\cite{mainz} collaborations).
Currently, the data points for the total cross section near the threshold
($E_\gamma \lesssim 1.2$ GeV),
where the $S$-wave $\omega N$ contribution is dominant,
are not enough for determining
$a_{\omega N}$ and $r_{\omega N}$
from the shape of the total cross section as a function of
the incident energy (excitation function)
through $\omega N$ rescattering in the final-state interaction.
We have measured ten data points of the total cross section
at incident photon
energies ranging from 1.09 to 1.15 GeV.
The $\omega$ meson mainly decays in the $\omega\to\pi^0\pi^+\pi^-$ mode
with a branching ratio of 89.2\%~\cite{pdg}.
It is, however, difficult to reproduce the background shapes in 
the $\pi^0\pi^+\pi^-$ invariant mass distributions measured with poor identification
for charged particles~\cite{hashimoto}.
Thus, we determined the cross sections using the $\omega\to \pi^0\gamma$ decay mode with a branching ratio of $8.40\%$.
In this letter, we present $a_{\omega N}$ and $r_{\omega N}$ extracted from the shape of the 
excitation function for the $\gamma p \to \omega p$ reaction.

A series of meson photoproduction experiments were conducted~\cite{exp}
using the FOREST detector~\cite{forest},
which was installed on the second photon beamline~\cite{tag2} at the Research Center for 
Electron Photon Science (ELPH), 
Tohoku University, Japan. In the present experiments, bremsstrahlung 
photons were produced from 1.2-GeV circulating electrons
in a synchrotron~\cite{stb} by inserting a carbon thread (radiator)~\cite{tag2}.
The photons collimated with two lead apertures of 10 and 25 mm in diameter
located 4.2- and 12.9-m downstream from the radiator, respectively, were 
incident on a 45-mm thick liquid-hydrogen target located 
at the center of FOREST.
The energies of the incident photons were analyzed up to 1.15 GeV by 
detecting the post-bremsstrahlung electrons with a photon-tagging counter, 
STB-Tagger II~\cite{tag2}.
FOREST consists of three different electromagnetic calorimeters (EMCs):
192 undoped CsI crystals, 
252 lead scintillating-fiber modules, and
62 lead glasses.
A plastic-scintillator hodoscope (PSH) is placed in front of each EMC
to identify charged particles.
FOREST covers a solid angle of $\sim{88\%}$ in total.
The typical photon-tagging rate was 20~MHz,
and the photon transmittance (the so-called tagging efficiency)
was $\sim${53\%}~\cite{tag2}.
The trigger condition of the data acquisition (DAQ),
which required for an event to have more than one final-state particles 
in coincidence with a photon-tagging signal~\cite{forest},
was the same as that in Ref.~\cite{dpipi-plb}.
The total number of collected events in DAQ was $1.79\times 10^9$.
The average trigger rate was 1.6~kHz, and the average DAQ efficiency was 80\%.

Event selection was made 
for the $\gamma{p}${$\to$}$\pi^0\gamma{p}${$\to$}$\gamma\gamma\gamma{p}$ reaction.
At first, events containing three neutral particles and a charged particle 
were selected.
The time difference between every 2 neutral EMC clusters
out of 3 was required to be less than thrice that
of the time resolution for the difference.
The two neutral EMC clusters giving the $\gamma\gamma$ 
invariant mass ranging from 50 to 220 MeV were selected, and 
the other EMC cluster was required to have an energy higher than 200 MeV. 
The charged particles were detected with the forward PSH.
Further selection was made by applying a kinematic fit with five constraints:
energy and three-momentum conservation,
and $\gamma\gamma$ invariant mass being the $\pi^0$ mass.
The momentum of the charged particle was obtained from the time delay
assuming that the charged particle had proton mass.
Events for which the $\chi^2$ probability was higher than 0.1 were selected.
When the number of combinations was more than 1 in an event,
the combination with the minimum $\chi^2$ was adopted.
Sideband-background subtraction was performed
for accidental-coincidence events detected in STB-Tagger II and FOREST.

\begin{figure}[b]
\begin{center}
\includegraphics[width=\figwidth]{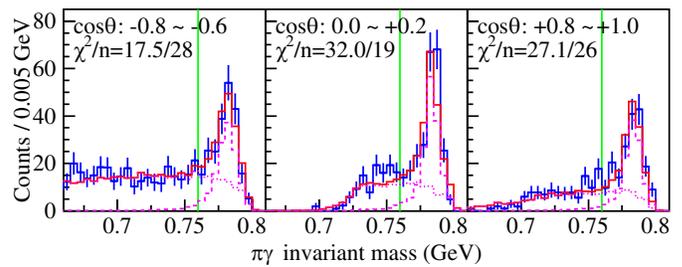}
\end{center}
\caption{Typical $M_{\pi\gamma}$ distributions for the highest incident energy 
group ($E_\gamma=1.144$--$1.149$ GeV). 
In each panel, the histogram (blue) shows the experimentally obtained 
$M_{\pi\gamma}$ distribution, and the solid curve (red) shows the sum of 
the $M_{\pi\gamma}$ distributions obtained in the simulation for the 
$\gamma p \to \omega p \to \pi^0\gamma p$ and 
$\gamma p \to \pi^0\pi^0 p$ reactions. 
The dashed (magenta) and dotted (cyan)
curves show these contributions. 
The angular region of $\omega$ emission in the $\gamma p$-CM frame 
is described in each panel.
The vertical lines show the lower limit 
$M_{\pi\gamma}=0.76$ GeV for selecting the $\omega$ produced events.
}\label{fig1}
\end{figure}

All the data for incident energies above 1.09 GeV
($E_\gamma=1.09$--1.15 GeV)
are divided into ten bins
(every bin includes 4 photon-tagging channels),
and 10 angular bins of the $\pi^0\gamma$ emission angle $\cos\theta$
in the $\gamma p$-CM frame.
The typical $\pi^0\gamma$ invariant mass ($M_{\pi\gamma}$) distributions are shown in Fig.~\ref{fig1}.
Each $M_{\pi\gamma}$ distribution shows a prominent peak 
with a centroid of $\sim 0.78$ GeV,
and has a broad background contribution in the lower side.
This background contribution is well reproduced by a Monte-Carlo (MC) simulation
based on Geant4~\cite{geant4}
for the $\gamma p \to \pi^0\pi^0 p\to \gamma\gamma\gamma\gamma p$ reaction,
where 1 $\gamma$ out of 4 is not detected with FOREST.
In the simulation, the five-fold differential cross sections
are assumed to be the same as those provided by the 2-PION-MAID 
calculation~\cite{2-pion-maid}.
The $M_{\pi\gamma}$ distributions for the
$\gamma p \to \pi^0\pi^0 p$ reaction are also plotted in Fig.~\ref{fig1}
where the same analysis is applied as for the $\gamma p \to \pi^0 \gamma p$ reaction.

\begin{figure}[b]
\begin{center}
\includegraphics[width=\figwidth]{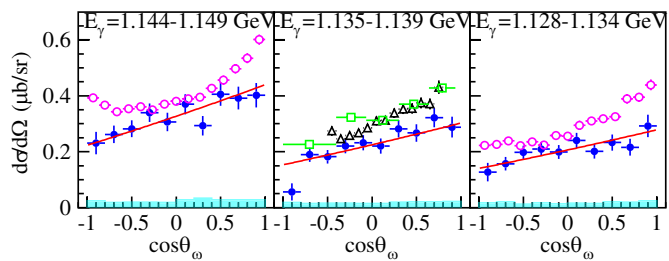}
\end{center}
\caption{Typical angular differential cross sections 
$d\sigma/d\Omega$ as a function of the $\omega$ emission angle $\cos\theta$
in the $\gamma p$-CM frame.
The range of the incident photon energies is described in each panel.
The filled circles (blue) represent the measured
$d\sigma/d\Omega$ in this work.
The shaded areas represent the systematic uncertainties of $d\sigma/d\Omega$s.
The solid curves show the fitted distribution with a $P$-wave contribution
of $\sigma^P_{\rm max}/5$ (see text).
The $d\sigma/d\Omega$ results from
SAPHIR~\cite{saphir}, CLAS~\cite{clas}, and A2~\cite{mainz} collaborations 
are depicted by open boxes (green), open triangles (black), and open circles (magenta), respectively.
The photon-energy coverages are 25, 18, 15, and $\sim 4.5$ MeV in 
SAPHIR, CLAS, A2, and our results, respectively.
}
\label{fig2}
\end{figure}

The $M_{\pi\gamma}$ distributions for 
the $\gamma p\to \omega p\to \pi^0\gamma p\to\gamma\gamma\gamma p$ 
and $\gamma p \to \pi^0\pi^0p \to \gamma\gamma\gamma\gamma p$ reactions in the MC simulation are fitted to the measured $M_{\pi\gamma}$ distribution 
for each emission-angle incident-energy bin 
only by changing the normalization coefficients.
Here, the events are generated according to the pure phase space
for the $\gamma p\to \omega p$ reaction.
The number of the $\omega$ produced events $N_\omega$ is estimated 
for $M_{\pi\gamma}\ge 0.76$ GeV after subtracting the background
$\gamma p \to \pi^0\pi^0p$ contribution for each bin.
The angular differential cross section is obtained
from $N_\omega (\cos\theta)$ as
\begin{equation}
\frac{d\sigma}{d\Omega}
= \frac{
N_\omega (\cos\theta)
}{
2\pi\Delta\cos\theta
N_\gamma
N_\tau \eta_{\rm acc} (\cos\theta)
{\rm BR}\left(\omega\to\gamma\gamma\gamma\right)
},
\end{equation}
with the incident photon flux including the DAQ efficiency correction $N_\gamma$,
the number of target protons $N_\tau$,
the multiplication of branching ratios 
for the $\omega\to \pi^0\gamma$ and $\pi^0\to\gamma\gamma$ decays
${\rm BR}(\omega\to\gamma\gamma\gamma)$,
and the detector acceptance calculated 
in the simulation $\eta_{\rm acc} (\cos\theta)$,
where $\Delta\cos\theta=0.2$.
Fig.~\ref{fig2} shows the typical $d\sigma/d\Omega$ distributions. 
The systematic uncertainty of $d\sigma/d\Omega$ is also given in Fig.~\ref{fig2}. 
It includes the uncertainty of event selection in the kinematic fit, 
that of counting $N_\omega$ due to the $M_{\pi\gamma}$ threshold,
that of acceptance owing to the uncertainties of the $d\sigma/d\Omega$
distributions for event generation in the simulation,
that of detection efficiency of protons,
and that of normalization resulting from $N_\tau$ and $N_\gamma$.

Every $d\sigma/d\Omega$ distribution shows a slight increase
with increase of $\cos\theta$.
A finite $P$-wave amplitude must produce asymmetric behavior of the 
angular distribution through the interference with the $S$-wave amplitude
although the $S$-wave contribution is expected to be dominant near the threshold.
The measured $d\sigma/d\Omega$s in this work are somewhat
lower than 
the world available data. 
The obtained $d\sigma/d\Omega$ depends on the incident-energy 
coverage because 
the cross section increases rapidly as the incident energy goes up.
The bin size of the incident energy is $\sim 4.5$ MeV in 
our results, while that for the
SAPHIR~\cite{saphir} data is 25 MeV, 18 MeV (CLAS~\cite{clas}),  and 15 MeV (A2~\cite{mainz}), respectively.
In Fig.~\ref{fig2}, 
the angular distribution 
obtained by the A2 collaboration
at $E_\gamma \simeq 1.14$ GeV
shows a shape being concave upward,
suggesting a
$P$-wave contribution, although any significant slope changes are not observed in
this work.
Apparently this deviation comes from the relative difference of centroid
photon-tagging energies by a few MeV. The uncertainty of the centroid 
photon-tagging energies is estimated to be 0.3\%, which corresponds to 3--4 MeV.
A calibration difference of photon-tagging energies needs to be incorporated
in the estimation of the systematic uncertainty for $a_{\omega p}$ and $r_{\omega p}$.

\begin{figure}[b]
\begin{center}
\includegraphics[width=\figwidth]{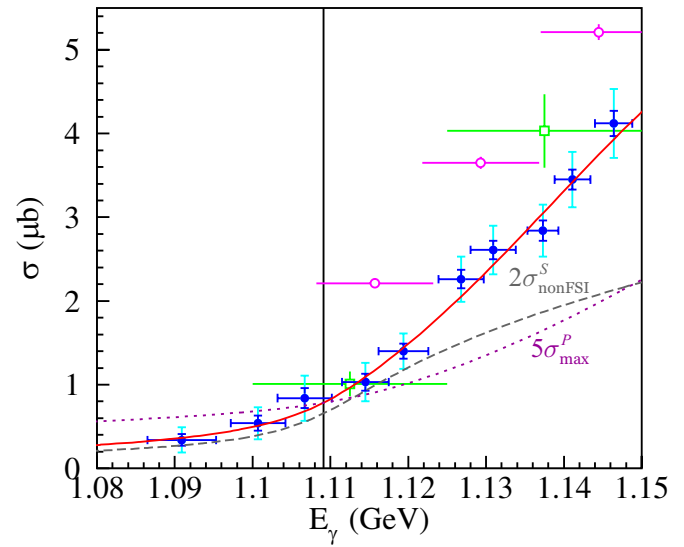}
\end{center}
\caption{Total cross section $\sigma$ as a function of $E_{\gamma}$. 
The filled circles (blue) represent the measured $\sigma$ in this work.
Each horizontal bar indicates the incident-energy coverage.
Each short vertical bar (blue) represents the statistical error of $\sigma$,
and the connected bars (cyan) indicate the upper- and lower-side systematic error
of $\sigma$. The vertical line (black) shows $E_\gamma$ corresponding to the 
reaction threshold for production of $\omega$ having a mass of the centroid.
The solid curve (red) shows the calculated excitation function with the 
parameters: $a_{\omega p}=-0.97+i0.07$ fm and $r_{\omega p}=+2.78-i0.01$ fm.
The dashed curves (gray) show $2\sigma^S_{\rm nonFSI}$ where
$\sigma^S_{\rm nonFSI}$ denotes the excitation function with
$a_{\omega p}= 0$ fm and $r_{\omega p}=$ 0 fm.
The dotted curve (purple) shows 
 $5\sigma_{\rm max}^P$ where
$\sigma_{\rm max}^P$ stands for 
the upper limit of the $P$-wave contribution.
The $\sigma$ results from
SAPHIR~\cite{saphir} and A2~\cite{mainz} collaborations 
are depicted by open boxes (green) and open circles (magenta), respectively.}
\label{fig3}
\end{figure}

The total cross section $\sigma$ is obtained by integrating $d\sigma/d\Omega$s
all over the ten emission-angle bins:
\begin{equation}
\sigma=\sum 2\pi\Delta\cos\theta \frac{d\sigma}{d\Omega}.
\end{equation}
Fig.~\ref{fig3} shows $\sigma$ as a function of the incident photon energy.
The excitation function
shows a monotonic increase, and finite yields
are observed below the threshold $E_\gamma^{\rm thr}$ for production of $\omega$ having 
the centroid mass.
The obtained cross sections show a systematic deviation from the world available data.
The uncertainty of tagging-energy determination (0.3\%) for the incident photon beam
may account for the deviation. 

We determine $a_{\omega p}$ and $r_{\omega p}$ from the shape of the 
excitation function.
We evaluate the excitation function 
for the $\gamma p \to \omega p$ reaction using a model with 
final-state $\omega p$ interaction (FSI) based 
on the Lippmann-Schwinger equation.
We assume 
that the $S$-wave contribution is dominant at $E_\gamma=1.09$--1.15 GeV.
The total cross section
for a fixed $\omega$ mass $M$ and $\gamma p$-CM energy $W$
can be calculated using a transition amplitude $T_{\gamma p\to\omega p}(W,M)$:
\begin{equation}
\sigma_0(W, M) = \frac{1}{16\pi W^2} \frac{p(W,M)}{k} \left|T_{\gamma p\to\omega p}(W,M)\right|^2,
\label{eq:cal}
\end{equation}
where $k$ and $p$ denote the momenta of an initial- and a final-state particles, respectively,
in the $\gamma p$-CM frame.
The total cross section $\sigma$ as a function of $E_\gamma$ is obtained by 
averaging $\sigma_0(W(E_\gamma),M)$ over available $\omega$ masses:
\begin{equation}
\sigma(E_\gamma) = \int^{W(E_\gamma)-m_p}_{m_{\pi^0}}\sigma_0(W(E_\gamma),M) L_\omega (M) dM,
\label{eq:cal2}
\end{equation}
where the probability $L_\omega (M)$ stands for a Breit-Wigner function with
a centroid of $M_\omega=782.65$ MeV
and a width of $\Gamma_\omega=8.49$ MeV~\cite{pdg}.

The $T_{\gamma p \to \omega p}$ is expressed by 
\begin{equation}
T_{\gamma p \to \omega p}
=
V_{\gamma p\to \omega p} + 
T_{\omega p \to \omega p} G_{\omega p\to\omega p} V_{\gamma p\to \omega p},
\end{equation}
where $T_{\omega p \to \omega p}$ stands for the $\omega p$ scattering 
amplitude, 
$G_{\omega p\to\omega p}$ denotes the $\omega p$ propagator, and
$V_{\gamma p \to \omega p}$ is the production amplitude without FSI.
We evaluate the matrix element for $T_{\gamma p \to \omega p}$
with on-shell approximations for $T_{\omega p \to \omega p}$ and
$V_{\gamma p \to \omega p}$,
and introduce a Gaussian form factor in the integration of $G_{\omega p\to\omega p}$.
This leads the matrix element of 
$T_{\gamma p \to \omega p}$
to the equation:
\begin{widetext}
\begin{equation}
\begin{array}{l@{}l}
&\left\langle
\displaystyle \vphantom{\vec{k}}
\omega p(p) \left|
\displaystyle \vphantom{\vec{k}}
T_{\gamma p \to \omega p}
\left| 
\displaystyle \vphantom{\vec{k}}
\gamma p(k)
\!\right.\right. \right\rangle \\
&=
\left\langle
\displaystyle \vphantom{\vec{k}}
\omega p(p) \left|
\displaystyle \vphantom{\vec{k}}
V_{\gamma p \to \omega p}
\left| 
\displaystyle \vphantom{\vec{k}}
\gamma p(k)
\!\right.\right. \right\rangle 
+\displaystyle \int \vphantom{\vec{k}}
\left\langle 
\displaystyle \vphantom{\vec{k}}
\omega p(p) \left|
\displaystyle \vphantom{\vec{k}}
T_{\omega p \to \omega p}
\left| 
\displaystyle \vphantom{\vec{k}}
\omega p(q)
\!
\right.\right. \right\rangle
{
\frac{\delta^3\left(\vec{q}-\vec{q}\,'\right)}{
W-H_0+i\epsilon
}
}
\left\langle 
\displaystyle \vphantom{\vec{k}}
\omega p(q') \left|
\displaystyle \vphantom{\vec{k}}
V_{\gamma p \to \omega p}
\left| 
\displaystyle \vphantom{\vec{k}}
\gamma p(k)
\!\right.\right. \right\rangle
d\vec{q}\,
d\vec{q}\,'\\
&\simeq
\displaystyle\left[
1+
8\pi\mu
\left\langle
\displaystyle \vphantom{\vec{k}}
\omega p(p) \left|
\displaystyle \vphantom{\vec{k}}
T_{\omega p \to \omega p}
\left| 
\displaystyle \vphantom{\vec{k}}
\omega p(p)
\!\right.\right. \right\rangle
\int \frac{d\vec{q}}{p^2-q^2+i\mu\Gamma_\omega }
\exp\left(-\frac{q^2}{\Lambda^2}\right)
\right]
\left\langle
\displaystyle \vphantom{\vec{k}}
\omega p(p) \left|
\displaystyle \vphantom{\vec{k}}
V_{\gamma p \to \omega p}
\left| 
\displaystyle \vphantom{\vec{k}}
\gamma p(k)
\!\right.\right. \right\rangle,
\end{array}
\end{equation}
\end{widetext}
where $H_0$ stands for the free Hamiltonian for the final-state $\omega p$,
and $\mu$ denotes a reduced mass between $\omega$ 
(with a mass of $M$) 
and the proton.
Here, we use a cut-off parameter $\Lambda=0.8$~GeV$/c$.
The $\left\langle
\displaystyle \vphantom{\vec{k}}
\omega p(p) \left|
\displaystyle \vphantom{\vec{k}}
T_{\omega p \to \omega p}
\left| 
\displaystyle \vphantom{\vec{k}}
\omega p(p)
\!\right.\right.\right\rangle $
is given by
$a_{\omega p}$ and $r_{\omega p}$:
\begin{widetext}
\begin{equation}
\left\langle
\displaystyle \vphantom{\vec{k}}
\omega p(p) \left|
\displaystyle \vphantom{\vec{k}}
T_{\omega p \to \omega p}
\left| 
\displaystyle \vphantom{\vec{k}}
\omega p(p)
\!\right.\right.\right\rangle
 = - \frac{1}{\left(2\pi\right)^2\mu}
\displaystyle
\left(\frac{1}{a_{\omega p}} + \frac{1}{2} r_{\omega p} p^2 - ip \right)^{-1}.
\end{equation}
\end{widetext}
The
$\left\langle
\displaystyle \vphantom{\vec{k}}
\omega p(p) \left|
\displaystyle \vphantom{\vec{k}}
V_{\gamma p \to \omega p}
\left| 
\displaystyle \vphantom{\vec{k}}
\gamma p(k)
\!\right.\right.\right\rangle$
is assumed to be a constant value of 1 in the incident-energy region of interest.

The dashed curve (gray) in Fig.~\ref{fig3} shows the excitation function
$\sigma^S_{\rm nonFSI}$
with $a_{\omega p}=0$ fm and $r_{\omega p}=0$ fm corresponding to non FSI condition,
which does not reproduce the experimental data.
FSI is necessary and the optimal set of $a_{\omega p}$ and $r_{\omega p}$
are determined 
to reproduce the experimentally obtained obtained cross section data.
The $\chi^2$ corresponding to the reproducibility is defined as
\begin{equation}
\chi^2 = \sum_{i=1}^{10} \frac{
\left(
\sigma_i - \alpha Y_i \right)^2
}{
\left(\delta\sigma_i^{(\rm stat)}\right)^2+
\left(\delta\sigma_i^{(\rm syst)}\right)^2
},
\end{equation}
where $\sigma_i$,
$\delta\sigma_i^{(\rm stat)}$,
$\delta\sigma_i^{(\rm syst)}$, and 
$Y_i$
denote 
the measured total cross section, its statistical error,
its systematic error,
and the yield estimated in Eq.~(\ref{eq:cal2}) 
by taking the coverage of incident energies into account,
respectively, for the $i$-th incident-energy bin.
The coefficient $\alpha$ for the overall normalization is determined to
minimize $\chi^2$ for each parameter set.
The deduced values are 
$a_{\omega p} = \left(-0.97^{+0.16}_{-0.16}{}^{+0.03}_{-0.00}\right)+
i \left(0.07^{+0.15}_{-0.14}{}^{+0.17}_{-0.09}\right)$
fm and $r_{\omega p}=\left(+2.78^{+0.68}_{-0.54}{}^{+0.11}_{-0.13}
\right)+i\left(-0.01^{+0.46}_{-0.50}{}^{+0.07}_{-0.00}\right)$ fm.
The first and second errors for each parameter
refer to the statistical and systematic uncertainties,
respectively.
The systematic uncertainty is estimated 
from that of the mean incident energy 
($\pm 0.3\%$) for each photon-tagging bin.
The solid (red) curve in Fig.~\ref{fig3} shows the excitation function with 
the optimal parameters.
No significant changes are observed for the $a_{\omega p}$ and $r _{\omega p}$ parameters
when we shift the incident photon energies by $\pm 0.3\%$. 
This is because these parameters are primarily determined by the shape
of the excitation function.

\begin{table*}[htb]
\caption{Deduced scattering parameters $a_{\omega p}$ and $r_{\omega p}$ for
several conditions. The second, third, fourth lines show the results for different 
$\Lambda$ cut-off parameters. The fifth line corresponds to the result with
taking the $P$-wave contribution into account where ${\rm Im\ } r_{\omega p}$ is fixed at 0 fm.
The sixth line represents the result with the assumption that
the energy dependence of the $\omega$ production amplitude 
is express by a single $N^*$ resonance as an extreme case.
}\label{tbl1}
\begin{center}
\renewcommand{\arraystretch}{1.35}
\begin{tabular}{ccccc}
\hline
parameters
 & Re $a_{\omega p}$ (fm)
 & Im $a_{\omega p}$ (fm)
 & Re $r_{\omega p}$ (fm)
 & Im $r_{\omega p}$ (fm) \\
\hline
$\Lambda=0.8 {\rm\ GeV}/c$ &
$\displaystyle -0.97^{+0.16}_{-0.16}{}^{+0.03}_{-0.00}$ &
$\displaystyle +0.07^{+0.15}_{-0.14}{}^{+0.17}_{-0.09}$ &
$\displaystyle +2.78^{+0.68}_{-0.54}{}^{+0.11}_{-0.13}$ &
$\displaystyle -0.01^{+0.46}_{-0.50}{}^{+0.07}_{-0.00}$ \\
$\Lambda=0.6 {\rm\ GeV}/c$ &
$\displaystyle -1.11^{+0.14}_{-0.16}{}^{+0.03}_{-0.02}$ &
$\displaystyle +0.12^{+0.17}_{-0.17}{}^{+0.15}_{-0.11}$ &
$\displaystyle +2.78^{+0.81}_{-0.57}{}^{+0.04}_{-0.11}$ &
$\displaystyle -0.00^{+0.44}_{-0.57}{}^{+0.11}_{-0.13}$ \\
$\Lambda=1.0 {\rm\ GeV}/c$ &
$\displaystyle -0.89^{+0.16}_{-0.18}{}^{+0.01}_{-0.00}$ &
$\displaystyle +0.04^{+0.14}_{-0.12}{}^{+0.13}_{-0.08}$ &
$\displaystyle +2.78^{+0.62}_{-0.51}{}^{+0.13}_{-0.09}$ &
$\displaystyle +0.01^{+0.47}_{-0.50}{}^{+0.03}_{-0.05}$ \\
\hline
$P$-wave contribution &
$\displaystyle -0.96^{+0.16}_{-0.16}{}^{+0.03}_{-0.01}$ &
$\displaystyle +0.10^{+0.14}_{-0.14}{}^{+0.14}_{-0.11}$ &
$\displaystyle +2.85^{+0.77}_{-0.53}{}^{+0.10}_{-0.15}$ & 
$\displaystyle 0.00$ \\
\hline
single $N^*$ contribution &
$-0.87^{+0.15}_{-0.12}{}^{+0.04}_{-0.02}$ & 
$+0.22^{+0.14}_{-0.12}{}^{+0.11}_{-0.11}$ &
$+2.69^{+0.62}_{-0.55}{}^{+0.06}_{-0.12}$ &
$-0.04^{+0.48}_{-0.69}{}^{+0.04}_{-0.14}$ \\
\hline
\end{tabular}
\end{center}
\end{table*}

The parameters may 
be somewhat affected by the adopted $\Lambda$.
We also determine $a_{\omega p}$ and $r_{\omega p}$
for $\Lambda=0.6$ and 1.0 GeV/$c$.
The obtained values are summarized in Table~\ref{tbl1}.
Although $\left|{\rm Re}\, a_{\omega p}\right|$ 
and $\left|{\rm Im}\, a_{\omega p}\right|$ become larger with
decrease of $\Lambda$,
changes of $a_{\omega p}$ and $r_{\omega p}$ 
are not significant among the realistic $\Lambda$ values.

The asymmetric behavior of the angular distribution mainly comes from
interference between $S$- and $P$-wave contributions.
The dotted curve (magenta) in Fig.~\ref{fig3} shows the shape of the 
$P$-wave excitation function 
where $\sigma_0^{P}(W,M)\propto p^3/k$ is assumed.
The finite width of $\omega$ makes the $P$-wave excitation function rather flat,
and the $P$-wave contribution does not explain the gap 
at higher incident energies between the data and calculation without FSI.
We also fit the excitation function adding a $P$-wave contribution
to the experimental data by fixing ${\rm Im}\, r_{\omega p}=0$~fm,
obtaining the values given in  Table~\ref{tbl1}.
The optimal coefficient to the $P$-wave contribution is 0, 
and the $P$-wave total cross section is 0 with an error of $\sigma^P_{\rm max}$.
The dotted curve in Fig.~\ref{fig3} 
corresponds to $5\sigma^P_{\rm max}$. 
The asymmetric behavior in the angular distribution shown in Fig.~\ref{fig2}
requires a finite $P$-wave contribution.
The solid curve in Fig.~\ref{fig2} corresponds to a solution under the 
condition that the $P$-wave contribution in $\sigma$ is 
$\sigma^P_{\rm max}/5$. We can conclude that 
the $P$-wave contribution in $\sigma$ is negligibly small
in determination of $a_{\omega p}$ and $r_{\omega p}$.

We have assumed that $V_{\gamma p \to \omega p}$ is constant
since the coverage of incident energies is narrow
($E_\gamma=1.09$--1.15 GeV) for several overlapping $N^*$s with a very wide width.
We deduce the scattering parameters with ${\rm Im}\, r_{\omega p}=0$~fm
by assuming a single $N^*$ contribution $D_{13}(1700)$ as an extreme condition:
\begin{equation}
V_{\gamma p \to \omega p}\propto
\left(W^2-M_{N^*}^2+ i M_{N^*}\Gamma_{N^*}\right)^{-1}
\end{equation}
where $M_{N^*}=1.7$ GeV and $\Gamma_{N^*}=0.2$ GeV~\cite{pdg}.
The change of each parameter from the constant $V_{\gamma p \to \omega p}$
is not significant.

\begin{figure}[htbp]
\begin{center}
\includegraphics[width=\figwidth]{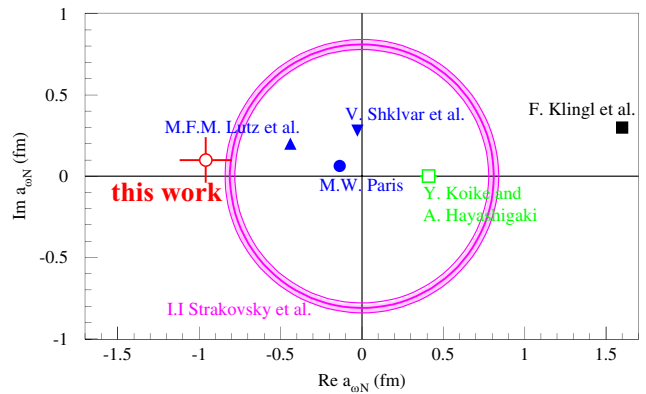}
\end{center}
\caption{Real and imaginary parts of spin-averaged $a_{\omega N}$
obtained in this work (red). 
The donut (magenta) represents the experimentally obtained 
$\left|a_{\omega N}\right|$ using a VMD model~\cite{mainz}.
Other markers indicate $a_{\omega N}$s obtained in the theoretical 
works on 
an effective Lagrangian approach~\cite{omega1} (black),
a QCD sum-rule analysis~\cite{omega2} (green),
coupled-channel analyses~\cite{omega3,omega4,omega5} (blue).
}
\label{fig4}
\end{figure}

Fig.~\ref{fig4} shows the real and imaginary parts of 
1/2 and 3/2 spin-averaged $a_{\omega p}$ obtained 
by assuming a constant $V_{\gamma p \to \omega p}$
in this work
together with the previously obtained values.
It is consistent with 
 $\left|a_{\omega p}\right|=0.82\pm 0.03$ fm given
by the A2 collaboration~\cite{mainz}. 
The other values correspond to the theoretical predictions.
The positive ${\rm Re}\, a_{\omega p}$ value,
giving an attraction,
is rejected at a confidence level higher than $99.9\%$.
The repulsion is found to be much stronger than the $\pi N$ ones,
and no bound or virtual state is expected for $\omega N$.
Slightly attractive $\omega$-nucleus ($\omega A$) interactions are reported with potential depths 
at normal nuclear density
of $-42\pm 17\pm 20$ MeV~\cite{a-metag} and $-15\pm 35\pm 20$ MeV~\cite{a-cbelsa}
from $\omega$ photoproduction from nuclei.
The measurement of $\omega$ line shape 
shows a decrease of  $\omega$ mass by $9.2\%\pm 0.2\%$ (corresponding to
$\omega A$ attraction)
without any in-medium broadening~\cite{ozawa,naruki}.
The relation between strong $\omega N$ repulsion and $\omega A$ attraction
would be a subject of future discussions taking into consideration
spin-dependent terms, higher partial waves, and partial restoration of chiral symmetry.

In summary, the total cross sections have been measured
for the $\gamma p \to \omega p$ reaction near the threshold. 
The $\omega$ is identified
through the $\omega\to \pi^0\gamma$ decay.
The spin-averaged scattering length $a_{\omega p}$ 
and effective range $r_{\omega p}$ between the $\omega$ and proton are
estimated from the excitation function
at incident photon energies ranging from 1.09 to 1.15 GeV:
$a_{\omega p} = \left(-0.97^{+0.16}_{-0.16}{}^{+0.03}_{-0.00}\right)+
i \left(0.07^{+0.15}_{-0.14}{}^{+0.17}_{-0.09}\right)$
fm and $r_{\omega p}=\left(+2.78^{+0.68}_{-0.54}{}^{+0.11}_{-0.13}
\right)+i\left(-0.01^{+0.46}_{-0.50}{}^{+0.07}_{-0.00}\right)$ fm.
The real and imaginary parts for $a_{\omega p}$ and $r_{\omega p}$
are determined separately for the first time.
A small $P$-wave contribution does not affect the obtained values.
The positive ${\rm Re}\, a_{\omega p}$ value indicates repulsion.

\begin{acknowledgments}
The authors express gratitude to the ELPH accelerator staff for stable operation 
of the accelerators in the FOREST experiments.
They acknowledge Mr.\ Kazue~Matsuda, Mr.~Ken'ichi~Nanbu, and Mr.~Ikuro~Nagasawa for their technical assistance in the FOREST experiments.
They received help at the early stage of this work from Dr.~Hiroyuki~Kamano.
They also thank Prof.~Igor~I.~Strakovsky for providing all the available 
numerical values of cross sections for the $\gamma p \to \omega p$ reaction.
They are grateful to Prof.\ Mark~W.~Paris for giving us the numerical values 
on the total cross sections of a single partial wave.
One of the authors (TI) expresses heartfelt gratitude to Dr.~Shuntaro~Sakai for 
several useful conversations.
This work was supported in part by the Ministry of Education, Culture, Sports, Science 
and Technology, Japan (MEXT) and Japan Society for the Promotion of Science (JSPS)
through Grants-in-Aid 
for Specially Promoted Research No.\ 19002003,
for Scientific Research (A) Nos.\ 24244022 and  16H02188,
for Scientific Research (B) Nos.\ 17340063 and 19H01902,
for Scientific Research (C) No.\ 26400287, and
for Scientific Research on Innovative Areas Nos.\ 18H05407 and 19H05141.
\end{acknowledgments}


\begin{thebibliography}{00}
\bibitem{machleidt}
R.~Machleidt, Adv.\ Nucl.\ Phys.\ {\bf 19}, 189 (1989).
\bibitem{sxn1}H.~Shen, H.~Toki, K.~Oyamatsu, K.~Sumiyoshi, Nucl.\ Phys.\ 
A {\bf 637}, 435 (1998).
\bibitem{sxn2}
B.~Abbott {\it et al.}, Phys.\ Rev.\ Lett.\ {\bf 119}, 161101 (2017).
\bibitem{mainz}
I.I.~Strakovsky {\it et al.} (A2 collaboration at MAMI), Phys.\ Rev.\ C {\bf 91}, 045207 (2015).
\bibitem{omega1}
F.~Klingl, T.~Waas, and W.~Weise, Nucl.\ Phys.\ A {\bf 650}, 299 (1999).
\bibitem{omega2}
Y.~Koike, and A.~Hayashigaki, Prog.\ Theor.\ Phys.\ {\bf 98}, 631 (1997).
\bibitem{omega3}
M.F.M.~Lutz, Gy.~Wolf, and B.~Friman, Nucl.\ Phys.\ A {\bf 706}, 431 (2002);\\
 ibid.~{\bf 765}, 495 (2006).
\bibitem{omega4}
V.~Shklyar, H.~Lenske, U.~Mosel, and G.~Penner,
Phys.\ Rev.\ C {\bf 71}, 055206 (2005).
\bibitem{omega5}
M.W.~Paris, Phys.\ Rev.\ C {\bf 79}, 025208 (2009).
\bibitem{saphir} J.~Barth {\it et al.}
(SAPHIR collaboration), 
Eur.\ Phys.\ J.\ A {\bf 18}, 117 (2003).
\bibitem{clas} M.~Williams {\it et al.} (CLAS collaboration),
Phys.\ Rev.\ C {\bf 80}, 065208 (2009);
ibid.\ {\bf 80}, 065209 (2009).
\bibitem{cbelsa} A.~Wilson {\it et al.} (CBELSA/TAPS collaboration),
Phys.\ Lett.\ B {\bf 749}, 407 (2015).
\bibitem{pdg}
M.~Tanabashi {\it et al.} (Particle Data Group), Phys.\ Rev.\ D {\bf 98}, 030001 (2018).
\bibitem{hashimoto}
R.~Hashimoto {\it et al.}, Few-Body Sys.\ {\bf 54}, 1135 (2013).
\bibitem{exp}
T.~Ishikawa {\it et al.}, JPS Conf.\ Proc.\ {\bf 10}, 031001 (2016).
\bibitem{forest}
T.~Ishikawa {\it et al.}, Nucl.\ Instrum.\ Meth.\ A {\bf 832}, 108 (2016).
\bibitem{tag2} T.~Ishikawa {\it et al.}, Nucl.\ Instrum.\ Meth.\ A {\bf 622}, 1 (2010);\\
T.~Ishikawa {\it et al.}, Nucl.\ Instrum.\ Meth.\ A {\bf 811}, 124 (2016);\\
Y.~Matsumura {\it et al.}, Nucl.\ Instrum.\ Meth.\ A {\bf 902}, 103 (2018);\\
Y.~Obara {\it et al.}, Nucl.\ Instrum.\ Meth.\ A {\bf 922}, 108 (2019).\\
\bibitem{stb}
F.~Hinode {\it et al.}, Proc.\ of 2005 Particle Accelerator Conference, 2458 (2005).
\bibitem{dpipi-plb}
T.~Ishikawa {\it et al.}, Phys.\ Lett.\ B {\bf 772}, 398 (2017);\\
T.~Ishikawa {\it et al.}, Phys.\ Lett.\ B {\bf 789}, 413 (2019).
\bibitem{geant4}
S.~Agostinelli {\it et al.}, Nucl.\ Instrum.\ Meth.\ A {\bf 506}, 250 (2003);\\
J.~Allison {\it et al.}, IEEE Trans.\ Nucl.\ Sci.\ {\bf 53}, 270 (2006);\\
Geant4 website $\langle${http://geant4.cern.ch/}$\rangle$.
\bibitem{2-pion-maid} 
A. Fix, H.~Arenh\"ovel, Euro.\ Phys.\ J.\ A {\bf 25}, 115 (2005);\\
2-PION-MAID website $\langle${https://maid.kph.uni-mainz.de/twopion/}$\rangle$.
\bibitem{a-metag}
V.~Metag {\it et al.}, Prog.\ Part.\ Nucl.\ Phys.\ {\bf 67}, 530 (2012);\\
V.~Metag, Hyperfine Interact.\ {bf 234}, 25 (2015).
\bibitem{a-cbelsa}
S.~Friedrich {\it et al.} (CBELSA/TAPS collaboration), Phys.\ Lett.\ B {\bf 736}, 26
 (2014).
\bibitem{ozawa} 
K.~Ozawa {\it et al.}, Phys.\ Rev.\ Lett.\ {\bf 86}, 5019 (2001).
\bibitem{naruki} 
M.~Naruki {\it et al.}, Phys.\ Rev.\ Lett.\ {\bf 96}, 092301 (2006).
\end{thebibliography}
\end{document}